C35

# 35

# THE GLOBALIZATION OF SCIENCE

## *The Increasing Power of Individual Scientists*

MAREK KWIEK

C35S1

## Introduction: The Emergent Global Science

C35P1

At the country level, science consists of two distinctive and heterogeneous systems: the global science and national science systems (Marginson & Xu, 2021). National science systems have become embedded in global science and countries, albeit for different reasons, but mostly to increase their economic competitiveness and to do everything they can to harness global knowledge to national economic needs. However, accessing and using the riches of global knowledge can occur only through scientists. Consequently, the research power of nations, among other factors, relies on the research power of individual scientists—their capacity to collaborate internationally and to tap into the global networked science is key. Being beyond global science networks and working on purely local research agendas, the academic community risks marginalization, thereby causing a loss of the interest among their national research-subsidizing patrons as well as losing the opportunity to influence the development of science.

C35P2

Global networked science can be analyzed through a variety of methodologies; however, quantitative science studies are probably best equipped to explore the extent of the globalization of science in spatial and temporal, individual and collective, national and cross-national dimensions using global publication and citation data. The global changes in how science is conducted are fundamental, and the accounts of these transformations abound (Adams, 2013; Gui et al., 2019; Wagner, 2008; Wang & Barabàsi, 2021). The general picture is well-known: for example, as Dong et al. (2017) show in their study of science in the past 100 years, the size of a publication's author list tripled and the rate of international collaborations increased 25 times; moreover, over 90% of the world-leading innovations (as measured by the top 1% most-cited papers) generated by teams in the 2000s was four times higher than that in the 1900s. The number of scholars and the number of publications grew at an exponential rate, doubling every 11 and 12 years, respectively. Finally, the share of single-authored publications shrank from 80% to 15%, with science shifting from individual work to collaborative effort.

C35P3 Further, the global map of science has changed in the past 100 years, with the increasing global diversification of scientific efforts—from the absolute dominance of the northeastern United States, the United Kingdom, and Germany in the 1900s to the leadership of both US coasts and Continental Europe in the second half of the 20th century to the rapid rise of research in Asia and other continents in the 21st century (Dong et al., 2017, p. 1444). The global science system currently indicates a larger, more competitive multicentric core. In terms of social network analysis, a bipolar world of science led by Anglo-Saxon countries is gradually being replaced by a tripolar world, which includes Europe, North America, and Asia-Pacific.

C35P4 Consequently, what has emerged in the past three decades is "a truly global scientific system" (Melkers & Kiopa, 2010, p. 389) or "a multipolar science world" (Veugelers, 2010) in which the scientific workforce is differently located, new trends in international collaboration have emerged, and the distribution of publication impact between traditional science powerhouses and the new entrants differs from decade to decade. Science is increasingly becoming a global system that comprises both advanced and less developed countries, with the global connectedness in science becoming important for both (Barnard et al., 2015). The depth and breadth of global science intensify, and the size of the global science network increases. The globalization of science implies a growing number of countries participating in international research collaboration and the ties between countries being much closer than before, thereby leading to decentralization (Gui et al., 2019) or pluralization (Marginson, 2018) of science. Collaboration remains dominated by science superpowers such as the United States, the United Kingdom, Germany, and several European countries, but countries where science is still emerging—such as China, followed by Brazil and South Korea—are ever more influential in the global network of science. The traditional Anglo-American academic hegemony is being challenged by new entrants (Marginson & Xu, 2021) in an increasing number of academic fields.

C35P5 Collaboration processes in science occur within different geographical units and, therefore, can be classified as regionalization, nationalization, and globalization; however, publication and citation data indicate that we are moving toward "a truly interconnected global science system" (Waltman et al., 2011, p. 574) in which globalization intensifies more than the other two processes. Using distance-based measurements of globalization, Waltman et al. (2011) reveal an evolution from a loosely connected 20th-century nation state science system to a 21st-century interconnected and internationally networked global science system, characterized by increasingly large distances among research partners. Science is globalizing at a steady rate; the authors have calculated what they termed the mean geographical collaboration distance for science as a whole, showing that between 1980 and 2009, the distance increased from 334 km to 1,553 km. The increase in collaboration distances occurred at different speeds: for example, the proportion of rather long partnerships (publications with the geographical collaboration distance of more than 5,000 km) has increased almost fivefold (Waltman et al. 2011, p. 576).

C35P6 The emergent picture of global science differs substantially from the traditional perspectives of how science works and which basic layers it consists of; specifically, the global networked science that challenges the traditional accounts of relationships between science and nation states (Kwiek, 2005) and welfare states (see Mattei, 2009). We have studied the changing relationships between the university and the state under globalization pressures; however, our main focus was on the impact of globalization on public sector services, welfare state architectures and funding, viewing higher education as an important claimant to public financing and analyzing higher education as directly competing with other segments of the welfare state (Kwiek, 2005, 2015) rather than on the globalization of science itself.

C35P7 From a global perspective, the most important factor in the gradual development of studies on the globalization of science was probably the increasing availability of digital data on scholarly inputs and outputs—the data on research funding, productivity, and collaboration, paper citations, and academic mobility—that offer unprecedented opportunities to explore the structure and evolution of science (Fortunato et al., 2018). Without access to global data, it would have been impossible to study the global networks of scientists, institutions, and ideas, novelty in science, academic career dynamics, the role of team science, or the citation dynamics from a global perspective. The globalization of science is currently explored under different conceptual labels and research agendas: the science of science (Clauset et al., 2017; Fortunato et al., 2018; Wang & Barabàsi, 2021; Zeng et al. 2017), meta-research or research on research (Ioannidis, 2018), computational social science (Edelman et al., 2020), quantitative science studies and studies of science and technology and its indicators (Glänzel et al., 2020), and others. In the previous decade, there has been an influx of natural, computational, and social scientists who together "have developed big data-based capabilities for empirical analysis and generative modeling that capture the unfolding of science, its institutions, and its workforce" (Fortunato et al. 2018, p. 1). For example, the science of science complements contributions from related fields such as scientometrics, informetrics, economics of science, and sociology of science. Social science is believed to be entering a golden age, with a rise in interdisciplinary teams working together that are leveraging the explosive growth of available data and computational power, as part of the big data revolution (Buyalskaya et al., 2021). In other words, the globalization-driven big data revolution in science is utilized to study the globalization of science itself.

C35S2

# Global Science and Nation States

C35P8

Generally, in the past 400 years, science has been affected by two major currents: nationalization and denationalization, with the latter often referred to as "globalization" (Crawford et al., 1993). At various levels, one or the other trend dominated in science. The primary reason why the nationalization trend is powerful despite globalizing pressures is that higher education, labor markets, science career paths, knowledge-producing institutions, and research funding are overwhelmingly national. Consequently, global science has a strong national relevance and all national science systems have at least some global relevance. There is no global science without a national funding base for research and training: Global science requires national funding to keep research infrastructure running and personnel costs covered. There are no global salaries in academic science yet (although the idea can refer to the corporate science originating from multinationals, as in the case of global pharmaceutical or computing industries and their publications). Simultaneously, as Freeman (2010, p. 393) argues, the globalization of scientific and engineering knowledge is "the most potent aspect of modern globalization."

C35P9

The relationship between science and the nation state has traditionally been strong, as nation states were the main patrons and sponsors of research. However, Caroline Wagner et al. suggest that the shift in science toward the global actually challenges the relationship between science and the state (Wagner et al., 2015, pp. 11–12). Since the end of the Cold War, the relationship between science funding and national identities as embodied in nation states has shifted considerably: The growth of international collaboration is decoupling science from the goals of national science policies (Wagner et al., 2015).

C35P10

Thus, the globalization of science theme captures the tension between global science and national sovereignty and can be viewed from the perspective of the sociology of science, particularly in the Mertonian tradition. Sociologists of science described four norms under which the scientific community works: universalism, disinterestedness, communalism, and organized skepticism (Merton, 1973). As portrayed in the historical sociology of science (Mallard & Paradeise, 2008), actual scientists were supposed to be intrinsically cosmopolitan figures: Mertonian norms were meant to present an accurate picture of the manner in which "science really works." Unlike politics, science was portrayed as disinterested and objective, and unlike religion, it was portrayed as skeptical. However, as the authors strongly emphasize, Robert Merton developed his ideas in the context of the Cold War in which the science practiced in the United States fundamentally differed from the science practiced in Soviet Russia and his ideas were first developed during World War II. Thus, it is worth remembering that the Mertonian tradition in the sociology of science, with its vision of ideal science and ideal scientists working in ideal meritocracy-based social environments and clear rules at the foundation of social stratification in science, is heavily embedded in a particular historical context (Kwiek, 2019a).

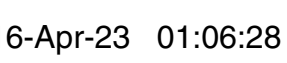

C35P11 In Merton's somehow ideal account, science is described as a curiosity-driven and disinterested systematic investigation, and its ultimate goal is to find truth without regard to political, social, or cultural interests (Cantwell & Grimm, 2018, p. 130). However, as the economics of science indicates, scientists and universities respond to incentives and even such shop-floor level characteristics of the science system as relative salaries in the sector—or entry academic salaries compared with entry salaries of other professionals—have an impact on who does science and who does not (Stephan, 2012, p. 5). Self-selection into science determines its future, as cross-sectoral mobility is rare and undervalued in most higher education systems.

C35P12 Recognition and reputation are key in science both as ends in themselves and as the means for acquiring the resources to continue doing science. Scientists are not rewarded for their efforts, like the time spent on research, but for their achievements—discoveries reported in publications, preferably with high impact in the scientific community and beyond. Stephan (2012) describes the nature of science not as a winner-take-all contest (in which there are no rewards for being second or third) but as a tournament arrangement (in which the losers obtain certain rewards as well which keeps individuals in the game of science despite not winning) (Stephan, 2012, p. 29). However, in terms of salaries, the top performers in research are clearly overrepresented among the academic top earners, at least in the 10 European systems studied (Kwiek, 2018a).

C35P13 Certain analysts emphasize the critical role of the global dimension in science, while others indicate that the national dimension—under changing national politics—may fight back. From the perspective of what Cantwell and Grimm term "the geopolitics of academic science," there are two prominent lines of competition between states: the competition for internationally mobile researchers and the competition to develop the strongest research universities. The world-class university project leads to the concentration of resources in selected elite universities and within certain disciplines, thereby possibly leading to the deprivation of public funds for other universities and other disciplines and possibly leading to the bifurcation of higher education systems between a small set of world-class elite institutions and a large set of demand-absorbing rest, thereby increasing vertical stratification in higher education and academic science (Cantwell & Marginson, 2018; Marginson, 2016). Academic science is reported, on the one hand, to be a global and cooperative enterprise and, on the other hand, to be a "nationalist endeavor designed to bolster state power relative to rivals" (Cantwell & Grimm, 2018, p. 144) with emergent tensions. In their account, we may now be entering a period of "cultural-economic nationalism, coupled with a technological-information globalism," with a constant tension "to reap the gains of global technology development for national purposes" (Cantwell & Grimm, 2018, p. 145).

C35P14 National geopolitics of higher education may go hand in hand with nationalism in academic science in which national interests and national purposes are of significance in the context of the arms race propelled by global university rankings. Scientific globalism has finally come to meet scientific nationalism today, but the two logics have coexisted for a long time, being rooted in the very idea of modern science—with the root metaphor of the former being the "republic of science" and for the latter being the "national innovation system." The rationale for support of science has been the addressing of grand scientific challenges and fostering international collaboration, on the one hand, and supporting global competitiveness and social and economic relevance, on the other (Sà & Sabzalieva, 2018, p. 153).

C35P15 In an influential paper on the emerging global model of the research university, Mohrman et al. (2008) argued that nation states have less influence over their universities than they did in the past. Global research universities have special missions which transcend the boundaries of the nation state, educate from a global perspective, and advance the frontiers of knowledge worldwide. Their special emphasis is on international interaction among universities across national boundaries. As the authors argue, these global research universities "operate beyond the control of the nation-state, leading to new policy dilemmas for national governments" (Mohrman et al., 2008, p. 15). Under the pressures of globalization, of which the globalization of science is a part, nation states are less able than before to control their destinies—they are more dependent upon universities for their knowledge production and their human capital, including doctoral students and doctorates in strategic research fields, both of which are essential for national, economic, and social development.

C35P16 Simon Marginson draws a useful distinction between "nation-centered" globalization (with an endless race between nations) and "world-system" globalization (which has a dynamic independence from nations and across all of them). The latter encourages not merely global convergence, but integration into a single system whose ultimate logic is the dissolution of the nation state. In science, the integration into a single system has already happened: Global science in practice "can no longer be wholly contained within a single country or blocked at the border. . . . States and WCUs [world-class universities] have to position themselves to advantage within these global systems that they can neither evade nor completely control" (Marginson, 2018, p. 73). World-class universities are among the most globalized social institutions today—while the national research environment and funding are of considerable significance. The tension is evident because research capacity is global but national funding for research and development (R&D) plays a key role in sustaining it. Therefore, higher education institutions, Marginson argues, are best understood as semi-dependent institutions that are irretrievably tied to the state; in contrast, world-class institutions are best understood as semi-independent institutions that are irretrievably tied to both the state and global science. Consequently, top institutions clearly have double allegiance: to nation states hosting (and still mostly funding) them and to global science with its strict rules and ranking-oriented definitions of success at institutional levels.

C35S3

## How Do Global Networks in Science Operate?

C35P17 The development of a global science system has its own dynamics of network formation. Research and scholarly inquiry are structured by rules, conventions, and intellectual property rights as well as by publishers' business agendas, on the one hand, and collegial academic gatekeeping, on the other (Marginson, 2018). Both national and global science are structured by the university hierarchy, and the knowledge produced in universities with prestige and resources has higher visibility and status than the knowledge produced elsewhere. There are also at least three other dimensions of inequalities: by country, by language, and by disciplines (Marginson, 2018, p. 36). Consequently, while global science is produced in most institutions, countries, languages, and disciplines, its highest impact is reserved for publications originating from world-class universities that are located mostly in Anglo-Saxon countries and published in English in science, technology, engineering, mathematics, and medicine (STEMM) disciplines.

C35P18 As Wagner et al. (2015) argue, "the active and robust global network is proof of its own usefulness. Researchers gain enough benefit from it that they are willing to extend the extra time and effort to maintain long-distance communications" (p. 12). The network is considered a new organization of science on the world stage: It adds to and complements national systems. The researchers examined a global network of science and have indicated that it has grown denser but not more clustered: There are a large number of additional connections, but they are "not grouping into exclusive 'cliques'" (Wagner et al., 2015, p. 1).

C35P19 The networks operate by clear rules. "They grow from the bottom up rather than from the top down. Networks become complex as they grow and evolve. Their organization is driven by the forces and structures—preferential attachment and cumulative advantage, trust and social capital creation, and the incentive system that leads scientists to share data and exchange information" (Wagner, 2008, p. 105). Perhaps what is most important for the future is that policymakers across the globe must first understand the dynamics of changes in order to be able to govern national science systems; it is only then that they will be able to devise incentives for scientists and integrate them skillfully within national recognition and reward systems in science. There is a long way to go from understanding global dynamics to incentivizing individual scientists within national systems so that what they do in science reflects at least a few national science policy priorities.

C35P20 The major issue is how to link academic knowledge production in one place with benefits resulting from this production to the same place as "the connection between supporting research and reaping its benefits can be quite tenuous" (Wagner, 2008, p. 107). The constantly evolving, bottom-up, autonomous, self-regulating, and self-focused nature of global science requires deep understanding and skillful support for certain directions of its development, for instance, toward more local applications, as compared to other directions. The reason for this is simple: Networks in science "cannot be controlled; they can only be guided." These networks evolve continuously according to the needs of scientists and the incentives made available to them. However, importantly, these needs and incentives most often "revolve around the desire for recognition in its broadest sense" (Wagner, 2008, p. 118). The best way to understand the dynamics of global science is to understand what drives academic scientists in their work, with the comprehension of mechanisms of academic recognition in the forefront. It is important to note that recognition in science is a rather fragile social and professional mechanism.

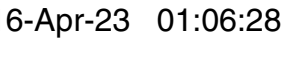

C35P21 In this chapter, we are particularly interested in the fundamental contrast between two opposing views: global science as a largely privately governed, networked, and normatively self-regulating institution (as in King, 2011) and global science as an emergent contributor to global collective public goods (as in Marginson, 2018). There is a discernible tension between the input side, or what motivates scientists to do science, and the output side, or what the results and outcomes of doing science are. As global science is increasingly outside the gaze of governments (King, 2011, p. 359), it may be moving to a more private sphere—"one of sociability rather than sovereignty, and the one that is characterized by loose ties and curiosity-driven scientific ambitions" (King, 2011, p. 359). The primary driver of global science is individual scientists who wish to collaborate with the best of their peers (Royal Society, 2011). Collaboration in research is curiosity-driven and reflects "the ambitions of individual scientists for reputation and recognition, not least as a means of pursuing their own research agendas," and new communication technologies facilitate the growing importance of "largely private" forms of global collaboration (King, 2011, p. 360). In other words, scientists may be increasingly collaborating as they wish, if they wish, and in the areas they wish, which, at a massive scale, is new from a historical perspective.

C35P22 Linking global science to national military and economic competitiveness, national economic policies, and science priorities is becoming increasingly difficult in the academic setting in which global science implies radically increasing individual freedom regarding the modalities and intensities of collaboration. The idea that science remains a powerfully state-driven and state-dependent rather than predominantly curiosity-driven and scientist-dependent is rather difficult to sustain. Global science is moving from scientific nationalism toward science as a public good, while simultaneously serving personal scientific ambitions of thousands of scientists and scholars.

C35P23 In King's account (King, 2011, pp. 362–367), self-regulatory and collaborative processes of science are conceptualized as networks that are beyond the supervision of governments. Global science is a constantly emergent system in the sense that it is the outcome of the numerous interdependent, individual, and decentralized normative decisions of individual scientists and scholars. Science is comprised of "interacting individuals and networks reproducing norms and standards"; these norms are principles for what is allowed and what is not, and the rules show which directions and procedures are desirable and which are not: "scientists form a moral community with an agreed outlook as to appropriate behavior" (King, 2011, p. 365). Clearly, governing this heterogeneous community and steering its academic behaviors, including collaboration and publishing behavior, is a tricky issue; however, with a thoughtful set of incentives, it is not impossible for national governments.

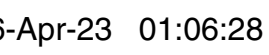

C35P24 What emerges through an accumulation of numerous decentralized and individual choices of scientists is convergence on the global research standards. King emphasizes that what is new in global science is that it occurs "largely behind the back of the nation-state, despite powerful political rhetoric espousing the competitive economic necessity of scientific nationalism in the knowledge economy" (King, 2011, p. 367). Understanding new dynamics in global science systems requires understanding the role of individual motivations for reputation and esteem in science: "science as a social institution always requires the energy and innovation that comes from ambitious and career-enhancing researchers" (King, 2011, p. 367).

C35P25 Collaboration in science often involves costs—that is, the time and the resources required as investments. Collaboration cannot be disentangled from reward systems in science, from how they operate, and what their major incentives are. In systems with powerful incentives to collaborate, collaboration grows faster; in systems with limited incentives to collaborate, collaboration grows at a slower pace (and new EU member states in Europe are a perfect example of systems of slow growths related to limited incentives in reward systems; see the EU-15/EU-13 comparison in Kwiek, 2020). Additional collaboration in science must be reflected in either the ways in which scientific reputation can be built up or in the ways in which competitive resources for research are nationally distributed, based on competing research proposals (Engels & Ruschenburg, 2008).

C35P26 Scientists—particularly those in the elite layers of affluent systems—appear to increasingly act as free agents, carefully selecting research collaborators in what Wagner terms the general shift from "national systems" to "networked science" and moving freely within a global network (Wagner, 2008, p. 25). According to Wagner, "national prestige is not the factor that motivates scientists as they work in their laboratory benches and computers. . . . within social networks, scientists seek recognition for their work and their ideas" (Wagner, 2008, p. 59). Precisely, emergent global science systems increasingly rely on King's "career-enhancing researchers" who seek recognition for what they do in science. If they cannot obtain such recognition in their national systems, they might choose to migrate to other systems or to quit academic science.

C35P27 The mechanisms of "cumulative inequality" in global science imply that the rich (in terms of reputation, citations, research funds, and personnel) get richer (King, 2011, p. 368); moreover, vertical stratification of the academic profession in global science creates a divide between the "haves" and "have-nots" (Wagner, 2008, p. 1; see my monograph on inequalities and the role of social stratification in science, Kwiek, 2019a). These new inequalities are compounded by the value ascribed to knowledge produced in different countries, disciplines, and in different languages, which are reflected in dominating citation patterns.

C35P28 As national ties in science weaken, the role of individual scientists and individual motivation appears to increase (Kato & Ando, 2016), and individual scientists compete intensely within an "economy of reputation," involving "battles over resources and priorities" (Whitley, 2000, p. 26). The growth of global science, among other factors, is an outcome of the rational choices of individual scientists seeking to maximize their own research output and impact (Hennemann & Liefner, 2015, p. 345). The phenomenon of preferential attachment—that is, "seeking to connect to someone already connected" (Wagner, 2018, p. 76)—guides scientists' collaboration behavior across systems and institutions. A scientist's rising reputation (and associated access to critical resources such as data, equipment, and funding) implies that "other researchers are increasingly likely to want to form a link with her" (Wagner, 2008, p. 61). Highly productive scientists attract similar individuals from elsewhere (King, 2011, p. 368) and international networks are created around these key people in global science, as they are highly attractive because they offer knowledge, resources, or both (Wagner, 2018, p. 70), while bearing in mind major gender difference: Male scientists are reported to be more internationally collaborative (and less collaborative in general) than female scientists (Kwiek & Roszka, 2020).

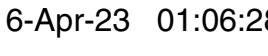

C35S4

## What Global Data Tell Us About the Globalization of Science

C35P29 In this section, we briefly describe the globalization of science using selected publication, collaboration, and citation data applied to several dimensions of globalization processes. The timeframe used is 2000–2020, unless otherwise stated, and the data come from Scopus (2021) and its SciVal (2021) functionality; the 25 countries (Top 25) analyzed are the largest global knowledge producers as of 2020 (articles only) and the 25 universities are top national knowledge producers (articles only) in the top 25 countries. The data were collected in the period of March 15–17, 2021.

C35S5

### The Globalization of Science Versus Institutions, Sectors, and Individuals

C35P30 Each scientist involved in academic knowledge production leaves traces of his/her activities in his/her printed publications; our knowledge regarding the globalization of science is generally based on numerous heterogeneous data sources (biographical, administrative, financial, publications, citations, collaboration etc.) produced at different levels (from the micro-level of individual scientists to the mezo-level of institutions to the macro-level of countries and regions) with different methodologies (from interviews to surveys to analyses of bibliometric data sets). However, the globalization of science can be traced using temporal, topical, geographical, and network analyses or traced over the years, countries, and institutions, research teams, and individual scientists, as well as academic disciplines by the expanding databases of globally indexed publications, with all commonly discussed limitations.

C35P31 The traces left by scientists in the form of globally indexed publications reveal the concentration of research at all levels, from individuals to institutions to countries. Among approximately 20,000 institutions active in the world (Scopus, 2021), there is no more than 1,000 involved in competitive, global academic knowledge production. The SciVal platform of the Scopus database (SciVal, 2021) indicates that in 2015–2020, the total number of academic institutions involved in global academic publishing was not higher than 9,000 (8,633). These were accompanied by institutions from corporate (6,130), government (2,523), medical (1,859), and other (797) sectors. In the period of analysis, the largest share of global knowledge production comes from the academic sector, followed by the government and corporate sectors. The top knowledge producer in the corporate sector is IBM, with Samsung, Microsoft, GlaxoSmithKline, and AstraZeneca in the top 10; the top 50 corporate institutions involved in global academic publishing include multinationals such as Pfizer, Intel, Merck, Siemens, Novartis, Johnson & Johnson, Airbus Group, Bayer, ABB Group, and Sanofi-Aventis. In the government sector, the top producer is the Chinese Academy of Sciences, with CNRS in France, Russian Academy of Sciences, National Research Council of Italy, and National Institutes of Health in the United States in the top 10; in the medical sector, the top producer is Mayo Clinic in Rochester, Minnesota, with Dana-Farber Cancer Institute in Boston, Massachusetts, in the top 10. Overall, from a global perspective, the academic sector is the key knowledge producing sector and a key participant in the globalization of science.

C35P32 If a threshold of 5,000 publications within the decade of 2010–2019 is used, then the number of all institutions above the threshold shrinks to 1,590, and these could be called world-class universities. There are 934 institutions with at least 10,000 publications, 153 with at least 50,000, and 24 with at least 100,000 publications of all types, globally. Harvard University is by far the largest global knowledge producer, with more publications than almost all countries (except for 22; for example, in Europe, Harvard has more publications than Denmark, Austria, Portugal, or Norway, as well as Mexico, Israel, or Malaysia globally).

C35P33 If we examine the research-focused rankings, the Leiden ranking 2020 lists 1,176 universities with at least 100 publications in the 2015–2018 period and the ARWU World University Ranking 2020 lists 1,000 universities. Specifically, in more regional terms, 41% of universities in the top 100 of the ARWU ranking are located in the United States; 66% of universities are located in one of the following five countries: the United States, the United Kingdom, France, Switzerland, and Australia; and the upper 10 countries constitute 83% of the locations.

C35P34 As globalization of research progresses, the concentration of research intensifies at the level of individual scientists and scholars with respect to both output and impact or publication and citation numbers. Four in ten of 6,167 Clarivate's highly cited researchers in 2020 originate from US universities (41.5%), seven in ten originate from the top five countries (71.8%), and 84.2% from the top ten countries. Only approximately 1% of globally publishing scientists (of approximately 15 million in the period 1996–2011) constitute the "continuously publishing core" of the academic profession, with at least a single paper published every year within the 16 years studied. However, they are responsible for 41.7% of all papers published in the same period (Ioannidis et al., 2014, p. 1). Moreover, approximately 1% of the most cited scientists in 118 scientific disciplines in 2015 received 21% of all citations, a sharp increase from 14% in 2000 (Nielsen & Andersen, 2021, p. 5). The upper 10% of scientists and scholars in terms of research productivity are responsible for approximately half of all academic knowledge production in 11 European systems across 7 major clusters of disciplines (and are often termed "research top performers") (see Kwiek, 2016, 2018b).

C35S6

## The Globalization of Science Versus Global Innovations

C35P35

While it is useful to focus on the overall potential of a country as viewed via its total number of publications, it is more revealing to trace global transformations through high-quality publications only. Specifically, in this section, we focus on the top 1% of highly cited publications (used as a proxy of high quality, with all limitations; see Tahamtan & Bornmann, 2019) and publications published in the top 1% of highly ranked journals. We assume that the top 1% of articles in terms of impact, as indicated through the citations attracted, are global innovations—or at least innovations globally recognized by other scientists—in academic science, and the publications in the upper 1% of journals are on average at least good candidates to become global innovations in the future.

C35P36

Table 35.1 presents the distribution of top publications in top knowledge-producing countries (as of 2020) in the two decades in the period 2000–2020 (country codes are provided in Table 35.4). The left panel indicates the changes in the percentages and the right panel in the numbers of publications over time. European systems—such as Switzerland, Belgium, and the Netherlands—from a global perspective, produce relatively high percentages and relatively small number of top publications. In terms of numbers, China already produces more top publications than the United States, and both are followed by the United Kingdom, Germany, Italy, and Australia. China continued to improve in terms of high-quality publications every year; in 2010, China had five times less of such publications than the United States; in 2015, it had only half of such publications compared with the United States; and in 2020, the difference increased substantially (with China overtaking the United States, with approximately 11,000 compared with approximately 8,000). All selected countries performed above expectations in their top publications, the expectation being the production of 1% of such publications; however, a few countries increased their numbers substantially: Apart from China, the highest increase in top publications in the past 5 years was noted in Italy (by 58%) in Europe as well as in Iran (by 348%) and India (by 174%) globally. Simultaneously, the number of top publications originating from the United States in 2020 and 2010 was similar, and a 17% decline was noted for the 2015–2020 period (Table 35.1, right panel); the numbers for other countries were only slightly declining or increasing.

C35T1

**Table 35.1 High-Impact Publications: Proportion (%) of Publications in the Top 1% of Publications by Citations**

| Country | Average 2000–2020 | 2000 | 2010 | 2015 | 2020 | Country | Total 2000–2020 | 2000 | 2010 | 2015 | 2020 |
|---|---|---|---|---|---|---|---|---|---|---|---|
| CHE | 2.9 | 2.1 | 3 | 3.5 | 2.4 | CHN | 67,497 | 107 | 1,561 | 4,550 | 10,900 |
| BEL | 2.3 | 1.2 | 2.3 | 2.8 | 2.3 | USA | 167,559 | 5,944 | 8,233 | 9,536 | 8,064 |
| AUS | 2.0 | 1.2 | 1.9 | 2.1 | 2.2 | GBR | 48,174 | 1,250 | 2,214 | 3,091 | 3,343 |
| NLD | 2.7 | 1.8 | 2.8 | 3 | 2.2 | DEU | 36,889 | 832 | 1,845 | 2,476 | 2,179 |
| GBR | 2.1 | 1.6 | 2.2 | 2.4 | 2.1 | ITA | 19,659 | 327 | 874 | 1,278 | 2,014 |
| ITA | 1.6 | 0.9 | 1.6 | 1.8 | 2.0 | AUS | 20,650 | 291 | 827 | 1,420 | 1,972 |
| SWE | 2.2 | 1.3 | 2.3 | 2.5 | 2.0 | CAN | 24,465 | 551 | 1,193 | 1,547 | 1,668 |
| CAN | 2.0 | 1.6 | 2.1 | 2.2 | 1.9 | IND | 9,000 | 62 | 266 | 559 | 1,529 |
| CHN | 1.2 | 0.2 | 0.7 | 1.2 | 1.8 | FRA | 23,919 | 565 | 1,151 | 1,535 | 1,511 |
| IRN | 0.8 | 0.1 | 0.4 | 0.6 | 1.8 | ESP | 15,373 | 194 | 715 | 1,068 | 1,311 |
| FRA | 1.7 | 1.1 | 1.7 | 1.9 | 1.7 | NLD | 18,538 | 358 | 923 | 1,231 | 1,128 |
| DEU | 1.8 | 1.2 | 2 | 2.1 | 1.6 | IRN | 4,655 | 2 | 78 | 246 | 1,101 |
| USA | 2.1 | 2.1 | 2.3 | 2.2 | 1.6 | KOR | 10,618 | 82 | 412 | 762 | 1,070 |
| ESP | 1.4 | 0.8 | 1.4 | 1.6 | 1.5 | JPN | 17,669 | 548 | 761 | 998 | 1,069 |
| TWN | 0.9 | 0.5 | 0.7 | 1 | 1.4 | CHE | 15,148 | 301 | 681 | 1,105 | 924 |

*Note*: Output in top 1% citation percentiles by country and publication year, 2000–2020, all publication types included, all fields of research and development combined, in descending order for 2020, top 15 countries in each panel only, in percent (left panel, world average = 1) and publication number (right panel).

C35P37

Publishing in top journals (Kwiek 2021) leads (on average) to higher field-normalized citation rates due to the very construction of journal percentile ranks in Scopus based on citations received in the previous 4 years. In Europe, the share of publications in high-impact journals that exceeds expectations is noted for Switzerland, the Netherlands, the United Kingdom, Belgium and Sweden, as well as for Australia, Canada, and the United States globally (all with over 4% of their publications in this category in 2020; see Table 35.2). Among East Asian countries, China, Korea, and Taiwan fare much worse. However, with regard to the number of publications in top journals, China is globally unbeatable in terms of increase in publication numbers, with 2,700 publications in 2010, 7,100 in 2015, and as many as 17,600 in 2020, which amounts to an increase of 149% in the 2015–2020 period, with very high probability of overtaking the United States in the next few years as it did in the case of high-impact publications. In certain fields of research, China in 2020 is already publishing a higher (in agricultural sciences and engineering and technologies) or equal (in natural sciences) number of articles in the top

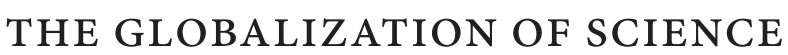

C35T2

**Table 35.2 Publications in High-Impact Journals, Proportion (%) of Publications in the Top 1% of Journals**

| Country | Average 2000–2020 | 2000 | 2010 | 2015 | 2020 | Country | Total 2000–2020 | 2000 | 2010 | 2015 | 2020 |
|---|---|---|---|---|---|---|---|---|---|---|---|
| CHE | 5.1 | 4.5 | 5.4 | 5.4 | 5.1 | USA | 339,080 | 1,1441 | 16,337 | 18,199 | 21,343 |
| NLD | 5.3 | 5.1 | 5.6 | 5.8 | 4.9 | CHN | 110,039 | 363 | 2,676 | 7,095 | 17,646 |
| AUS | 3.8 | 3.5 | 3.6 | 3.9 | 4.3 | GBR | 95,466 | 2,945 | 4,405 | 5,599 | 6,954 |
| CAN | 4.1 | 4.2 | 4.1 | 4.1 | 4.3 | DEU | 70,781 | 1,853 | 3,421 | 4,213 | 4,810 |
| GBR | 4.4 | 4.2 | 4.5 | 4.6 | 4.3 | CAN | 48,851 | 1,313 | 2,275 | 2,821 | 3,816 |
| USA | 4.5 | 4.6 | 4.7 | 4.4 | 4.3 | AUS | 38,068 | 725 | 1,502 | 2,545 | 3,730 |
| BEL | 4.4 | 4.0 | 4.8 | 4.7 | 4.2 | FRA | 47,307 | 1,343 | 2,400 | 2,813 | 2,874 |
| SWE | 4.3 | 3.4 | 4.2 | 4.9 | 4.2 | ITA | 35,611 | 965 | 1,666 | 2,152 | 2,515 |
| DEU | 3.6 | 3.0 | 3.9 | 3.7 | 3.6 | NLD | 35,891 | 920 | 1,748 | 2,318 | 2,482 |
| FRA | 3.5 | 3.2 | 3.7 | 3.6 | 3.2 | ESP | 31,612 | 531 | 1,549 | 2,090 | 2,385 |
| CHN | 2.0 | 0.9 | 1.2 | 1.9 | 2.9 | KOR | 24,742 | 245 | 1,001 | 1,892 | 2,301 |
| KOR | 2.7 | 1.8 | 2.4 | 2.9 | 2.9 | JPN | 38,464 | 1,792 | 1,712 | 1,856 | 1,981 |
| ESP | 3.0 | 2.4 | 3.2 | 3.3 | 2.7 | CHE | 25,368 | 563 | 1,189 | 1,632 | 1,961 |
| ITA | 3.0 | 3.2 | 3.2 | 3.1 | 2.6 | SWE | 20,362 | 492 | 861 | 1,380 | 1,497 |
| TWN | 2.7 | 2.5 | 2.8 | 2.9 | 2.3 | BEL | 16,297 | 371 | 800 | 1,065 | 1,172 |

*Note*: Publications in top 1% journal percentiles (by Scopus CiteScore percentile) by country and publication year, 2000–2020, all publication types included, all fields of research and development combined, in descending order for 2020, top 15 countries in each panel only, in percent (left panel, world average = 1) and publication number (right panel).

1% of journals than the United States. The largest remaining gap in article production in these journals by the two academic superpowers is in the medical sciences, as well as in humanities and social sciences, which are traditionally underrepresented in large data sets of the Scopus and Web of Science genre.

C35S7

## The Globalization of Science Versus Publishing Patterns in Academic Disciplines

C35P38

In general, research literature (usually focused on the STEMM fields) reveals that international collaboration has been on the rise across countries, institutions, and academic disciplines as well as among scientists and scholars. Social sciences and humanities are usually omitted from analyses, following arguments that neither Scopus nor Web of Science databases adequately reflect knowledge production in these fields. However, it is useful to show the changing distribution of the various collaboration types over time across all major fields of research and development, notwithstanding limitations. It is suffice to say that among 41,462 journals listed in Scopus, there are 5,002 journals allocated to arts and humanities and 10,199 allocated to the social sciences. Further, international research collaboration (and consequently global publishing patterns) can be analyzed in the context of three other collaboration types: institutional, national, and single authorship (or no collaboration). The four collaboration types are complementary, and the globalization of science can be analyzed through the changing intensity of international collaboration over time. The six fields of research and development used here, following the OECD, are agricultural sciences, engineering and technologies, humanities, natural sciences, medical sciences, and social sciences.

C35P39 Perhaps the most surprising effect of such a global disaggregated approach to academic publishing and collaboration patterns is the powerful and increasing gap between social sciences and the humanities. In the last two decades, while social sciences clearly follow the patterns characteristic of natural sciences, the humanities increasingly diverge from social sciences, moving in a fundamentally different direction in terms of the collaboration mix.

C35P40 Let us consider the collaboration mix for all fields of research and development combined (Figure 35.1) and compare this general picture with the picture of ongoing changes in collaboration in natural sciences, social sciences, and humanities over the period of two decades (2000–2020; Figures 35.2, 35.3 and 35.4). In our approach, the changing collaboration patterns studied through percentages at the country and institutional levels reflect the changing publishing patterns at the level of individual scientists and scholars affiliated with institutions in these countries. Thousands of individual-level publishing decisions are reflected in aggregated pictures of collaboration at higher levels of analysis.

C35P41 The collaboration patterns in natural sciences (Figure 35.2) follow the patterns of global science in general (or of all fields combined): in all countries, international collaboration has been on the rise in the last two decades. Increasing international collaboration has occurred at the expense of institutional collaboration and no collaboration (or single-authored) research, both of which have been declining in percentage terms; while institutional collaboration and solo research have been reducing, national collaboration has been stable or, in numerous cases, increasing in percentage terms. In particular, the stability of national collaboration both from a global perspective and in natural sciences indicates the importance of the national embeddedness of science. International collaboration does not appear to crowd-out national collaboration in any of the countries; at the global level (see World in Figures 35.1 and 35.2), national collaboration has increased substantially—from 26% in 2000 to 35% in 2010 and 42% in 2020.

C35P42 The striking feature of the changing collaboration mix by academic fields is that the role of international collaboration in the humanities is marginal, and in most countries it increases very slowly. In contrast, in the social sciences, the most important trend is the increase in international collaboration, predominantly at the expense of single-authored research—single-authored publications tend to dominate in the humanities;

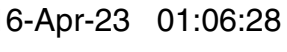

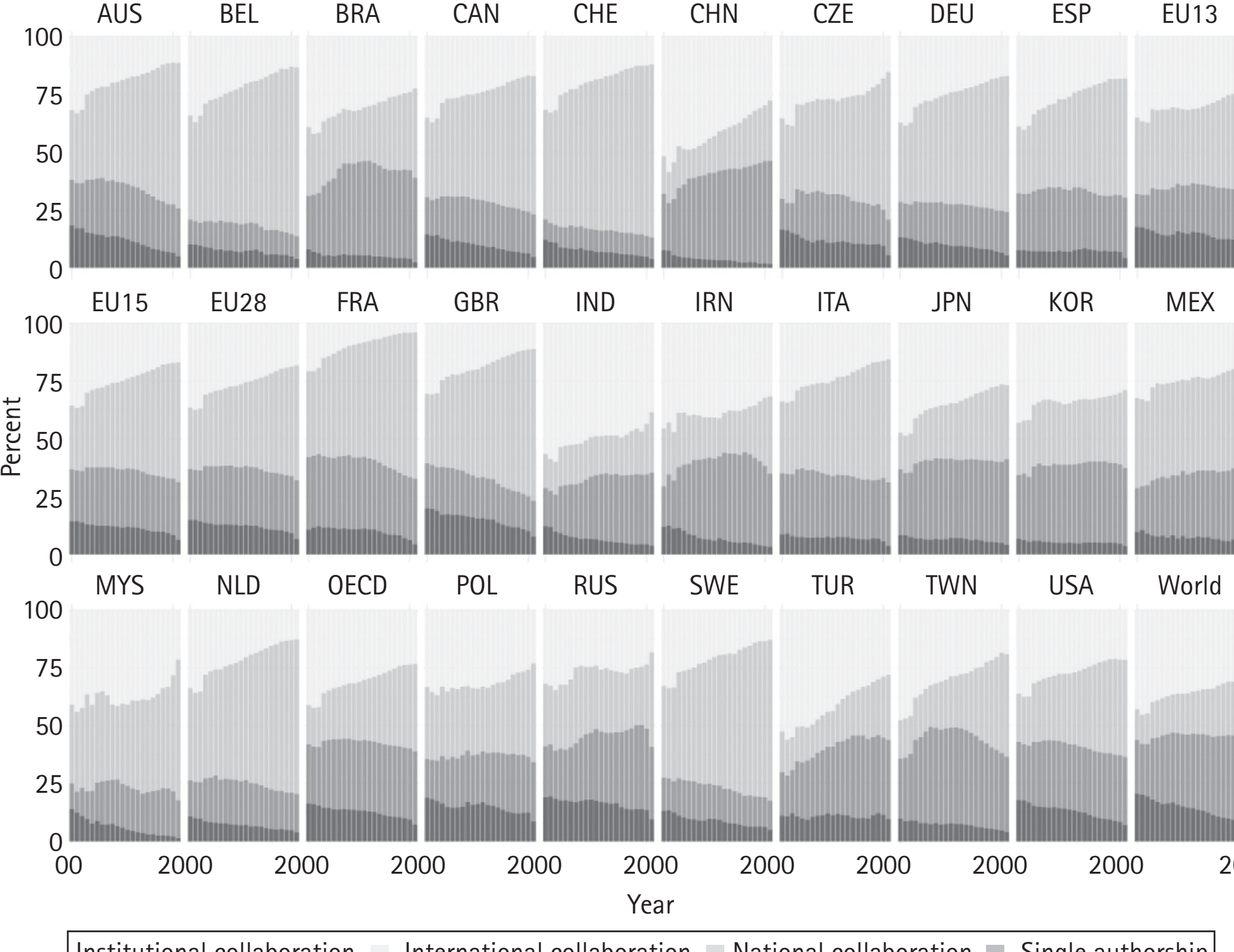

C35F1

**FIGURE 35.1** Collaboration (and publishing) patterns for all fields of research and development combined: powerful and increasing international collaboration at the expense of institutional collaboration, with stable national collaboration: top 25 global knowledge producers in 2020 (plus EU-28, EU-15, EU-13, OECD, and the world), articles only, SciVal data, 2000–2020 (%).

while in social sciences, the decrease in solo research is substantial (which is evident at the global level in the World box in Figure 35.3), the share of solo research in the humanities in almost all countries still exceeds 50%. Figures 35.3 and 35.4 graphically depict the powerful divergence, which appears to be increasing over time, between social sciences and humanities and has not been emphasized in current literature on the globalization of science.

C35P43

With regard to the powerful global social sciences/humanities divide, while the global percentage of single-authored articles declined from approximately half to approximately one-fourth (from 49% to 23% between 2000 and 2020) in the former, in the latter, there was only a slight decline from 67% to 56% at the global level (see World in Figure 35.4). In the social sciences, all 25 countries and 5 agglomerates of countries studied noted significant declines in shares of single-authored articles and, in most cases, significant increases in shares of internationally collaborative articles, with stable shares of national collaboration over time. International collaboration in the humanities has been relatively insignificant in most countries, except for several European systems. The share of solo articles in 2020 exceeded 40% of all academic knowledge production in the humanities in all countries and agglomerates studied, except for three European countries (Belgium, the Netherlands, and Switzerland) and four newcomers to the global top knowledge producers (China, Indonesia, Iran, and Malaysia). Further, single authorship is the dominating mode of publishing in the humanities and its share exceeds 50% in the most advanced economies: The percentage of solo articles in 2020 was 55% for EU-28, 55% for the OECD, and 51% for the United States.

FIGURE 35.2 Collaboration (and publishing) patterns in the natural sciences: powerful and increasing international collaboration at the expense of institutional collaboration, with stable national collaboration: top 25 global knowledge producers in 2020 (plus EU-28, EU-15, EU-13, OECD, and the world), articles only, SciVal data, 2000–2020 (%).

The changing publishing patterns have their implications for funding at the individual level and beyond. While in most national funding agencies and national excellence initiatives across the globe, social sciences and humanities are grouped together, it must be clear to the academic community, policymakers, and grant makers that the divergence in publishing patterns between the two academic fields has been widening in the last two decades.

Humanities are clearly non-collaborative, and clearly non-internationally collaborative, with powerful implications for metrics such as average output levels and average citation levels at the micro level of individual academics. Individual productivity in all fields except for the humanities is increasing mostly due to the full counting of

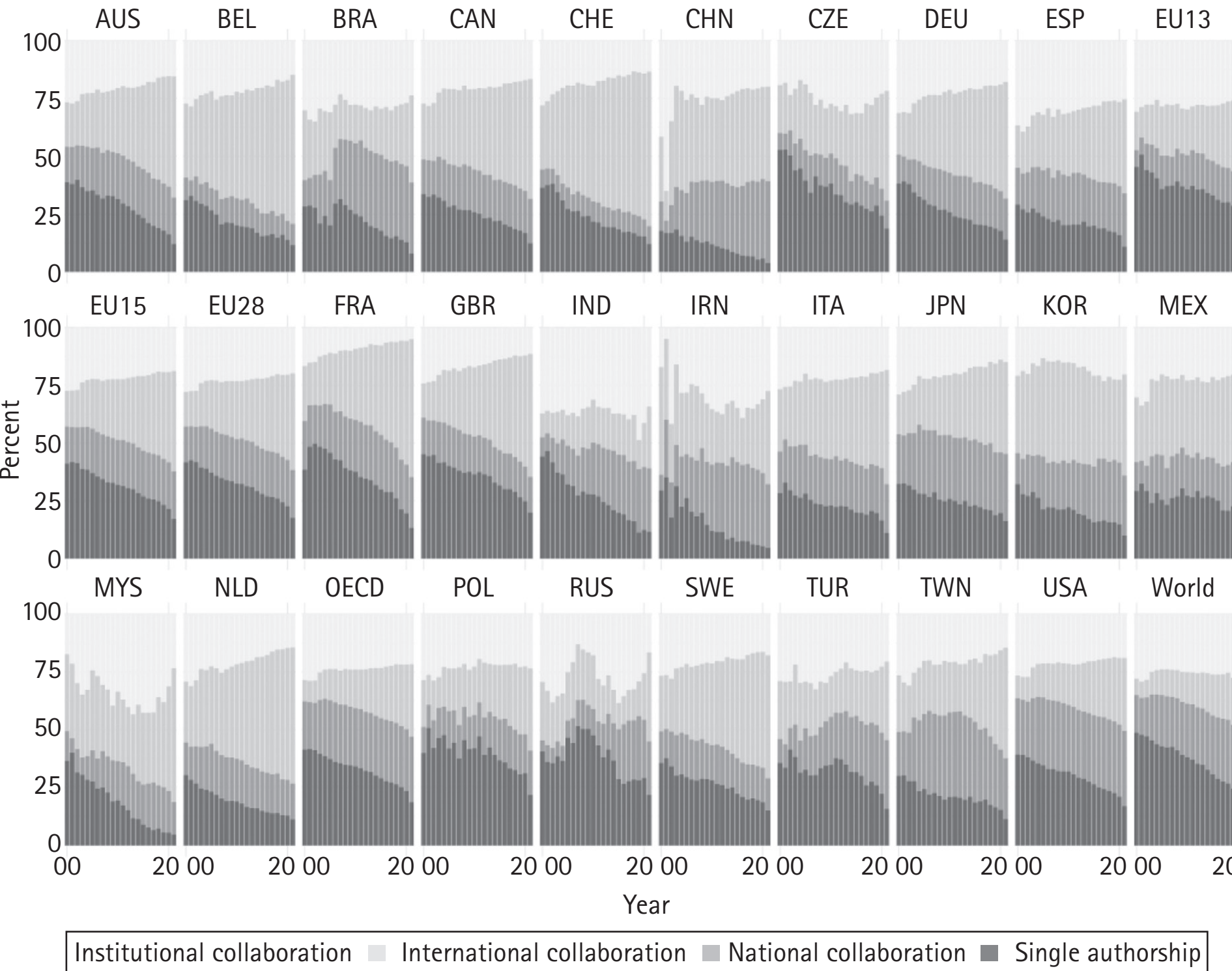

C35F3

**FIGURE 35.3** The collaboration (and publishing) patterns in the social sciences: increasing international collaboration with radically declining single authorship (no collaboration) and stable national collaboration: top 25 global knowledge producers in 2020 (plus EU-28, EU-15, EU-13, OECD, and the world), articles only, SciVal data, 2000–2020 (%).

publications written in teams; when the fractional counting method is applied, productivity is seen to be relatively stable over time. However, in the special case of the humanities, with single authorship as a dominating publishing pattern, individual output without using fractional counting methods, may appear small by comparison; moreover, as literature shows, citations to single-authored articles are lower than those to collaborative articles. The social sciences/humanities divide has its practical implications, disadvantaging humanists whenever they are in a head-on competition for research grants and awards with social scientists, and clearly promoting social scientists wherever the emphasis on publication and citation metrics dominates in the assessment of grant proposals. The traditional expression "social sciences and humanities" in the globalizing science and scholarship loses its traditional sense and can lead to unfair results in competitions among individuals, departments, and institutions.

C35P46 The changing international collaboration rate by discipline and country is presented in Figure 35.5: The top 25 countries can be clustered into low internationalization systems (such as Poland, Russia, Turkey, and India) and high internationalization systems (such as Switzerland, Sweden, Belgium, the United Kingdom in Europe, or Australia

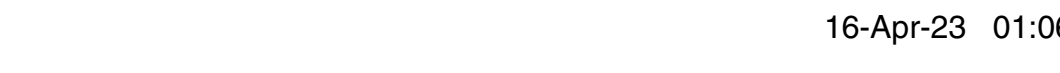

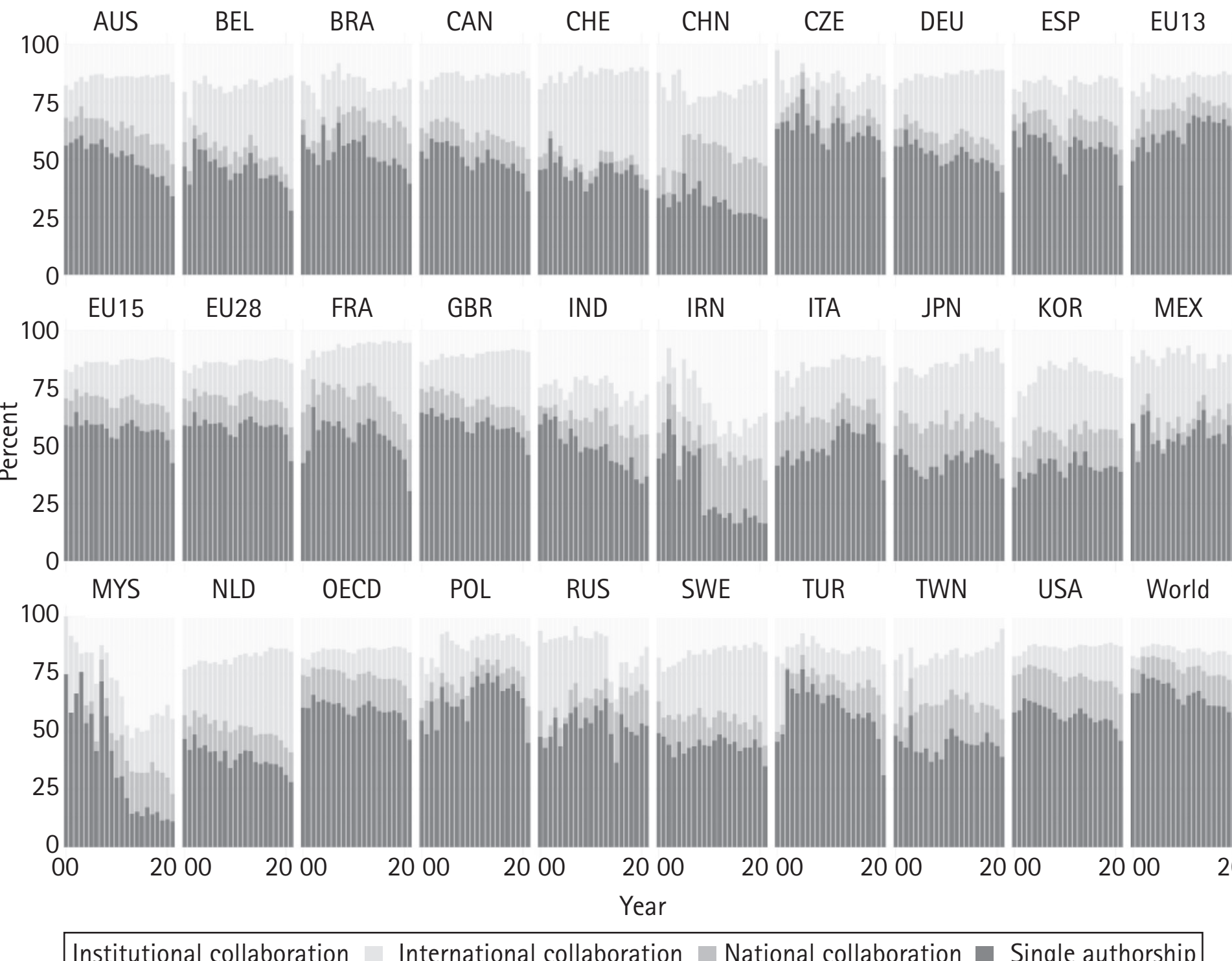

C35F4

FIGURE 35.4 Collaboration (and publishing) patterns in the humanities: powerfully dominating single-authorship (no collaboration), with a marginal role of slowly increasing international collaboration and stable national and institutional collaboration: top 25 global knowledge producers in 2020 (plus EU-28, EU-15, EU-13, OECD, and the world), articles only, SciVal data, 2000–2020 (%).

globally), with China in humanities and social sciences and the United States in agricultural sciences and natural sciences slowly increasing their international collaboration.

C35P47

The differences in the international collaboration rate in the top 25 countries are reflected only to a certain extent in the differences between the 25 largest knowledge producing universities (articles only) located within these countries (see Figure 35.6). For example, Harvard is more highly internationalized in terms of research than the United States, and Paris-Saclay University is more highly internationalized than France; the most highly internationalized among the selected national universities are the Swiss Federal Institute of Technology Zurich, Karolinska Institute, and KU Leuven, with rates of approximately 70% in 2020; the least internationalized are Anna University in India and Islamic Azad University in Iran (approximately 14% and 28% in 2020, respectively); in Central and Eastern Europe, both Lomonosov Moscow State University and Jagiellonian University in Cracow, with low and stable internationalization levels in 2000–2020 (approximately 30% and about 40% in the two decades), can be contrasted

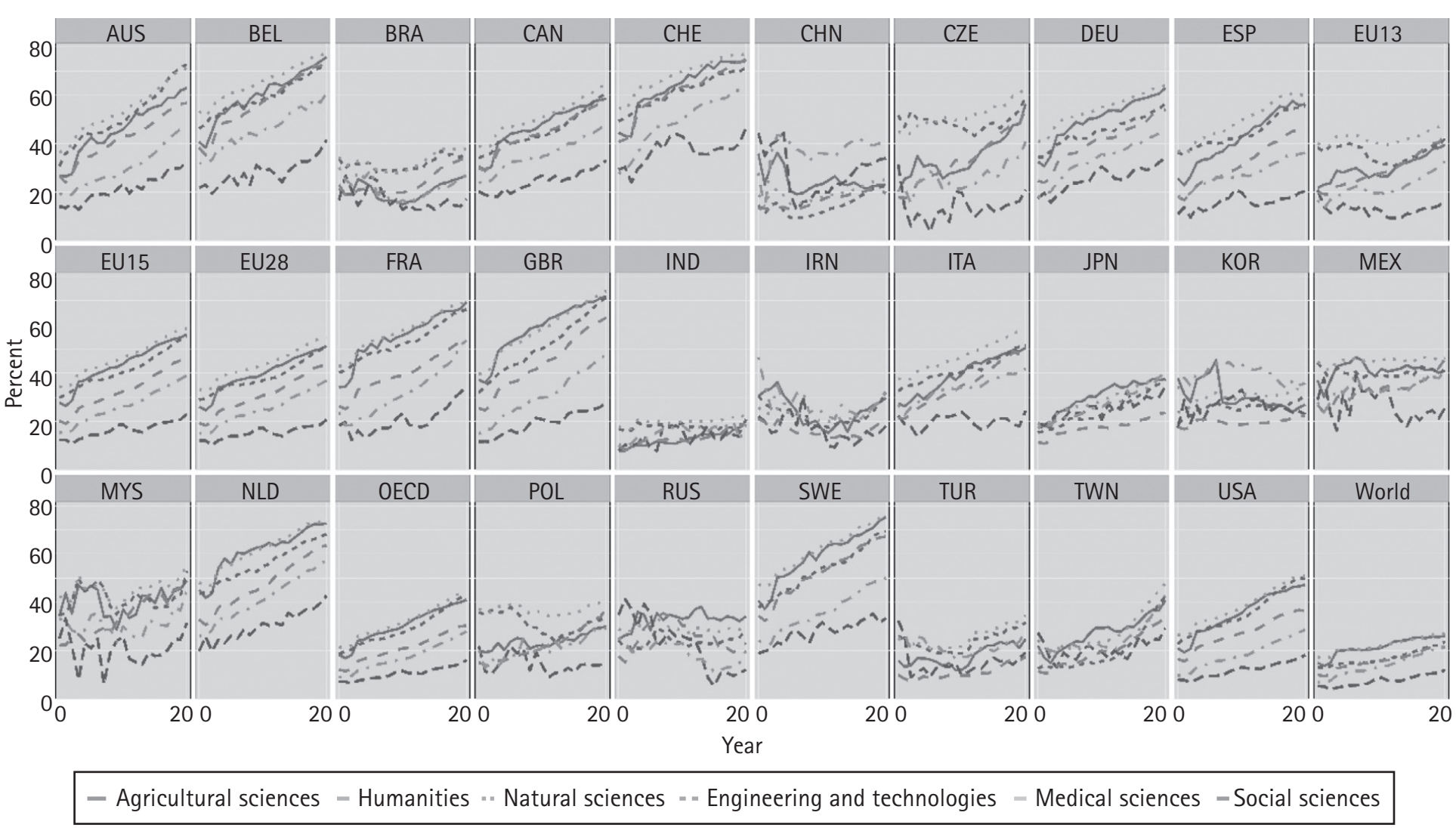

C35F5 **FIGURE 35.5** The international collaboration rate (percentage of internationally collaborative publications) by field of research and development, top 25 global knowledge producers in 2020 (plus EU-28, EU-15, EU-13, OECD, and the world), articles only, SciVal data, 2000–2020 (%).

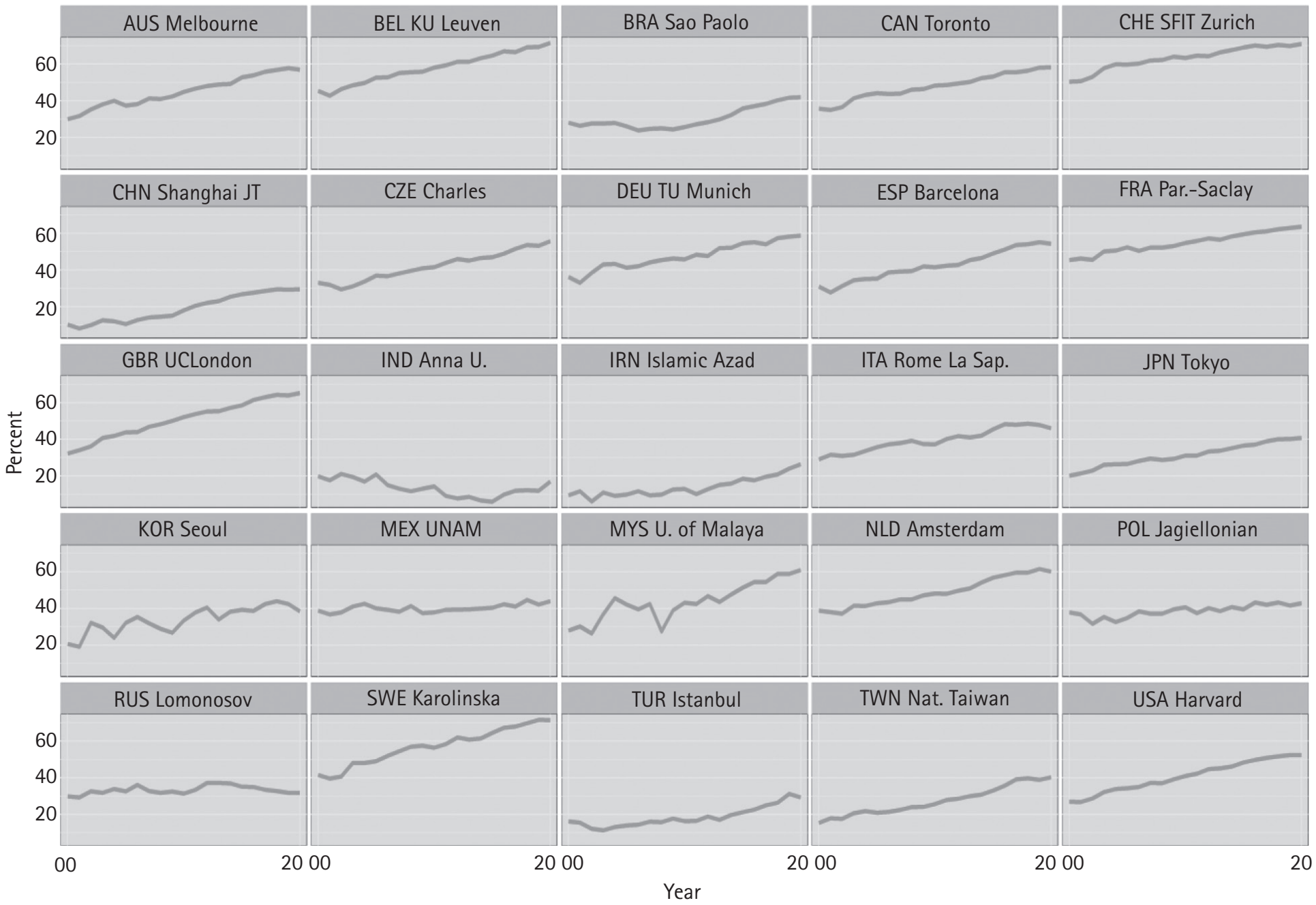

C35F6 **FIGURE 35.6** The international collaboration rate (percentage of internationally collaborative publications), all fields of research and development combined, 25 biggest knowledge-producing universities in the top 25 global knowledge-producing countries (as of 2020), articles only, SciVal data, 2000–2020 (%).

with Charles University in Prague, with the rate reaching approximately 55% in 2020 and increasing over time.

C35P48

Apart from changing percentages over time, the internationalization of science is also reflected in publication numbers changing over time. National output can be divided into two categories: articles involving international collaboration and all others—that is, domestic articles, including both single-authored and national and institutional collaborations (see Adams, 2013, p. 558). From this perspective, a major finding is that the increase in annual output in the period 2000–2020 in major European systems such as the United Kingdom, France, the Netherlands, Switzerland, Finland, Belgium, Sweden, and Germany and in non-European systems such as the United States, Australia, Canada, and Japan is almost entirely accounted for by international collaborations (see Figure 35.7). In contrast, in catching-up systems (such as India, Brazil, Iran, Mexico, Turkey, Russia, Poland, or Malaysia), there is an increase in national collaboration output. The most illustrative contrast is between the two global powerhouses: While the United States noted no increases in national publications, China noted a huge increase in the previous two decades (compare the two green areas for both countries in Figure 35.7). While domestic output in the former cluster of countries remained almost flat during the study period, the number of internationally coauthored articles increased steadily. The dark blue areas in Figure 35.7 indicate the growth in numbers of international collaborative publications, while the red line indicates the declining share of domestic publications; however, the declining share in a country does not have to imply declining numbers.

C35P49

The current power of research in the widely understood Western world resides in the growth of internationalization as seen through the volume of internationally coauthored output; the number of domestic publications has not changed in the past two decades. The globalization of science implies two different processes in two different system types: the growth of science in the Western world is almost entirely attributable to internationally coauthored publications, and its growth in the developing world is driven by both internationally coauthored and domestic publications, with different mixes in different systems.

C35S8

## The Globalization of Science Versus System Size, Citation Impact, and Preferred Collaboration Partner Countries

C35P50

The international collaboration rate across the 25 top countries is not generally correlated with national research output (defined as the total number of articles in 2000–2020). Plotting the percentage of internationally coauthored articles against system size in terms of article numbers (Figure 35.8) indicates that correlation is negligible ($R^2 = 0.03$). Bubble sizes confirm that systems with low international collaboration rates have low field-weighted citation impact (FWCI), as defined by Scopus, as in the case of Iran, Turkey, and India (as well as China, with the second-largest number

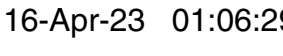

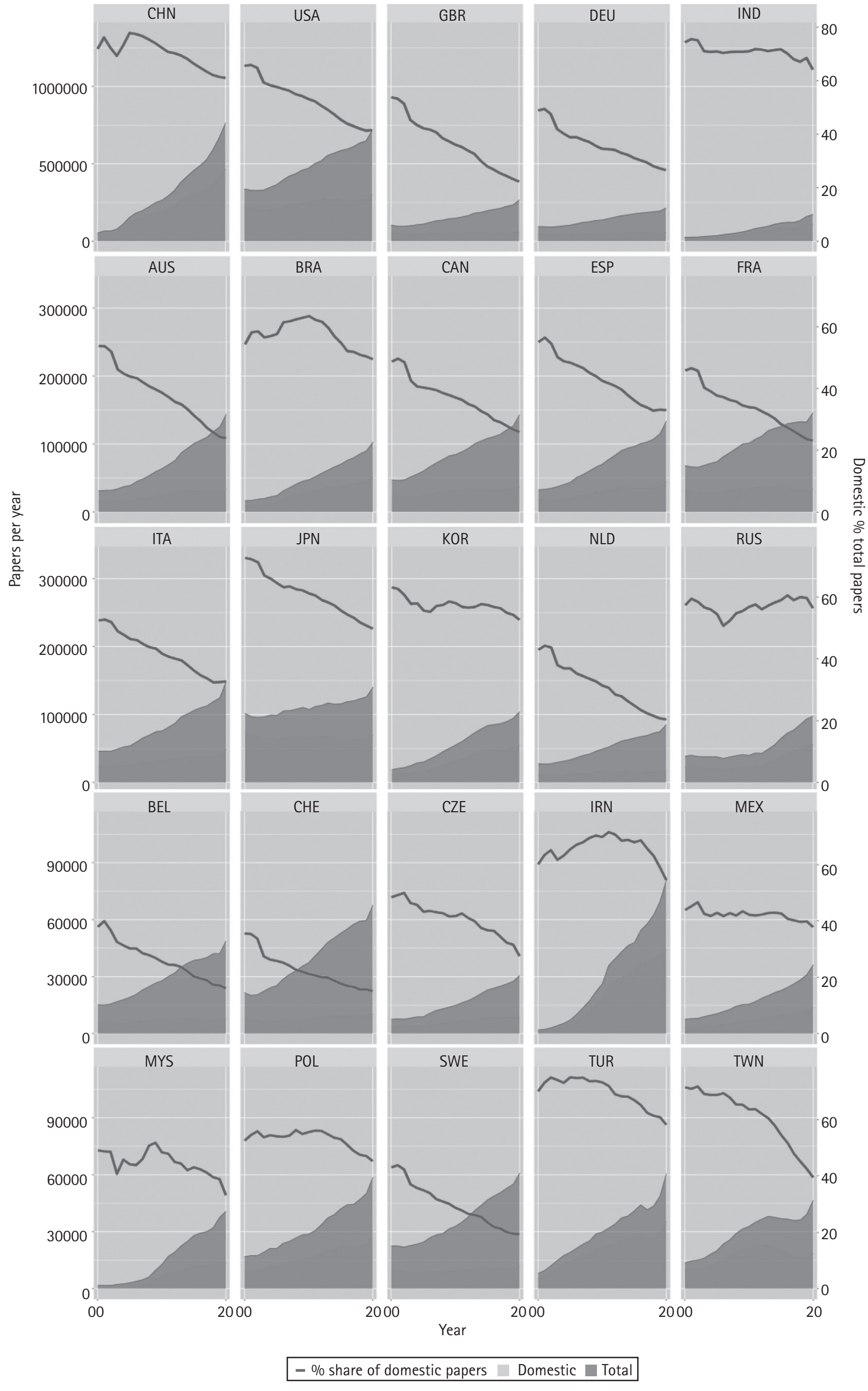

C35F7 **FIGURE 35.7** Total (dark blue), domestic (green), internationally collaborative publications (left axis) and the percentage of domestic publications (right axis) for the top 25 global knowledge producers (2000–2020).

C35F8 **FIGURE 35.8** Correlation between total national output 2000–2020 (articles only; log number) and percentage share of articles published in international collaboration, averaged for the period 2000–2020 (articles only); 95% confidence interval in gray; bubble size reflects average field-weighted citation impact (FWCI) for internationally coauthored articles for the period. All fields of research and development are combined.

of collaborative articles, which is a clear outlier in Figure 35.8, along with the United States).

C35P51 In Figure 35.9, the citation impact (the FWCI) of publications written in international collaboration is plotted against the citation impact of those written in national collaboration. Field normalization of scientometric indicators avoids distortions caused by different fields (Waltman & van Eck, 2019, p. 282). As measured in Scopus, the FWCI is the ratio of citations actually received to the expected world average (which equals 1) for the subject field, publication type, and publication year. Nationally coauthored publications are cited less often than expected in almost all European countries (i.e., countries to the left of the vertical line in Figure 35.9), with Brazil, Taiwan, Russia, China, and the United States slightly above the global average. Further, papers involving national collaboration had a higher citation impact on global science than international collaborations in the majority of countries (those below the red dashed line) for different reasons: the global superpowers China and the United States; Poland, France, and Iran, where both nationally and internationally coauthored papers had a high citation impact (cross-disciplinary differences are not discussed here because of word count constraints). At the aggregated level of all fields combined, the citation impact of internationally coauthored publications was above the expected field-weighted global average in the vast majority of European systems, but not global systems, analyzed. National collaboration produced globally impactful papers only in Spain, Italy, France, and Australia (quadrant 2) as well as in the United States and China (quadrant 4).

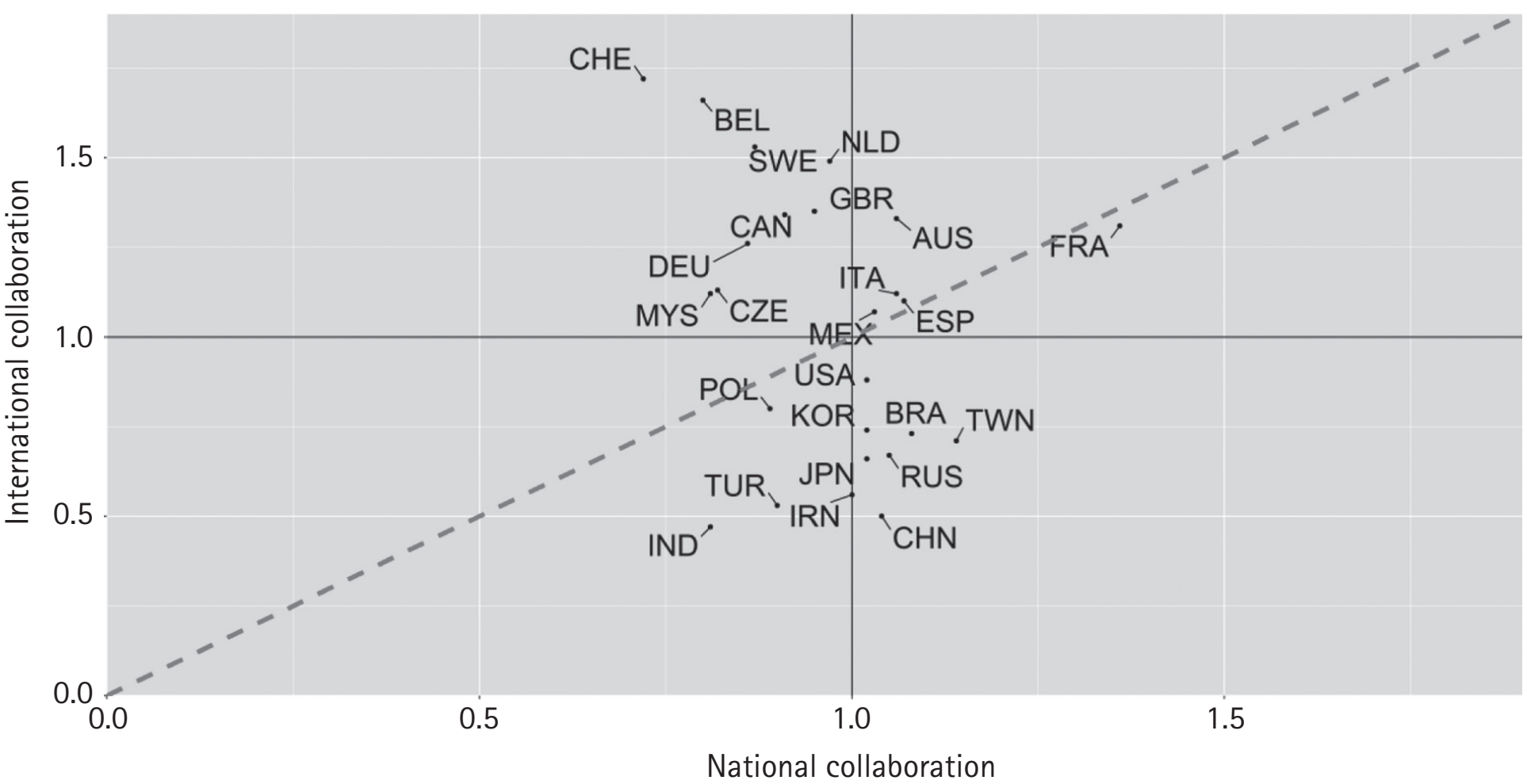

C35F9 **FIGURE 35.9** Field-weighted citation impact (FWCI) by publication type (internationally coauthored, nationally coauthored), articles only, self-citations included, average for the period 2000–2020, all fields of research and development combined.

C35P52 Finally, global science is characterized by different thickness of research pairings between countries and institutions: Certain pairings emerge as clearly preferred, following preferential attachment mechanisms in international collaboration. Preferred research pairings differ significantly in terms of their global visibility (as operationalized by the citation impact of internationally coauthored publications).

C35P53 We studied 25 thickest research pairings among our 25 top countries for the period 2015–2020, combined (Table 35.3, left panel). For all the countries involved, except the Netherlands, irrespective of the size of their science systems, the single most frequently collaborating partner is the United States. Other strong preferred collaboration links are intra-European or with China (in the case of the United Kingdom and Canada). The European integration in research, powerfully supported by European funding, enables the treatment of European countries as a single entity: In this case, among the globally thickest collaboration pairs, there would only be the United States (with Canada), China (with East Asian and Pacific systems of Japan, Korea, and Australia), and Europe. China and the United States form the most powerful global link, followed by the links between the United States and the United Kingdom, Germany, and Canada. Further, collaboration patterns for 28 European systems (Kwiek, 2020) indicate that geographical, linguistic, and historical ties still matter; for example, Spain is the top collaboration partner for Portugal, Finland for Estonia, Germany for Austria and the Czech Republic, France for Romania, and the Czech Republic for Slovakia. The United States remains the number-one collaborating partner for most European countries, including the largest knowledge producers (the United Kingdom, Germany, France, Italy, and Spain). However, in the top five ranks, the citation impact is highest for intra-European pairings of systems and European-American pairings; citation impact is lowest for

C35T3

**Table 35.3 Top 25 Collaboration Partnerships**

| Rank | Partner Country 1 | Partner Country 2 | Publications 2015–2020 | FWCI | Rank | Partner Country 1 | Partner Country 2 | Publications 2015–2020 | FWCI |
|---|---|---|---|---|---|---|---|---|---|
| 1 | USA | CHN | 344,409 | 1.93 | 1 | GBR | NLD | 63,171 | 3.25 |
| 2 | USA | GBR | 205,699 | 2.74 | 2 | USA | NLD | 71,185 | 3.22 |
| 3 | USA | DEU | 161,699 | 2.64 | 3 | USA | CHE | 65,749 | 3.12 |
| 4 | USA | CAN | 159,744 | 2.51 | 4 | GBR | FRA | 76,171 | 3.05 |
| 5 | GBR | DEU | 107,731 | 2.85 | 5 | ITA | DEU | 66,662 | 3.01 |
| 6 | USA | FRA | 106,311 | 2.85 | 6 | GBR | ESP | 60,658 | 3.01 |
| 7 | USA | AUS | 100,188 | 2.90 | 7 | GBR | AUS | 74,803 | 3.00 |
| 8 | USA | ITA | 99,589 | 2.83 | 8 | USA | ESP | 72,830 | 2.95 |
| 9 | GBR | CHN | 93,151 | 2.28 | 9 | GBR | ITA | 79,438 | 2.93 |
| 10 | CHN | AUS | 80,656 | 2.40 | 10 | DEU | FRA | 72,956 | 2.91 |
| 11 | GBR | ITA | 79,438 | 2.93 | 11 | ITA | FRA | 62,089 | 2.91 |
| 12 | USA | JPN | 78,246 | 2.40 | 12 | USA | AUS | 100,188 | 2.90 |
| 13 | GBR | FRA | 76,171 | 3.05 | 13 | GBR | DEU | 107,731 | 2.85 |
| 14 | GBR | AUS | 74,803 | 3.00 | 14 | USA | FRA | 106,311 | 2.85 |
| 15 | DEU | FRA | 72,956 | 2.91 | 15 | USA | ITA | 99,589 | 2.83 |
| 16 | USA | ESP | 72,830 | 2.95 | 16 | USA | GBR | 205,699 | 2.74 |
| 17 | USA | NLD | 71,185 | 3.22 | 17 | CHE | DEU | 62,336 | 2.68 |
| 18 | USA | KOR | 68,723 | 2.08 | 18 | USA | DEU | 161,699 | 2.64 |
| 19 | ITA | DEU | 66,662 | 3.01 | 19 | USA | CAN | 159,744 | 2.51 |
| 20 | USA | CHE | 65,749 | 3.12 | 20 | CHN | AUS | 80,656 | 2.40 |
| 21 | GBR | NLD | 63,171 | 3.25 | 21 | USA | JPN | 78,246 | 2.40 |
| 22 | CHE | DEU | 62,336 | 2.68 | 22 | GBR | CHN | 93,151 | 2.28 |
| 23 | ITA | FRA | 62,089 | 2.91 | 23 | CHN | CAN | 59,148 | 2.27 |
| 24 | GBR | ESP | 60,658 | 3.01 | 24 | USA | KOR | 68,723 | 2.08 |
| 25 | CHN | CAN | 59,148 | 2.27 | 25 | USA | CHN | 344,409 | 1.93 |

*Note*: Most prolific pairs 2015–2020, sorted by number of coauthored publications (left) and field-weighted citation impact (FWCI) of coauthored publications (right).

joint US-Chinese publications. Within these top five pairs, internationally coauthored papers are cited 3.01–3.25 times more than the world average for similar publications. The networks formed by the thickest collaboration links within the 25 top countries are depicted in Figure 35.10, based on frequency and citation impact.

## The Tensions of Global Science

The rise of new scientific powers as seen in the earlier empirical sections—in terms of collaborations, impact, and the role of highly innovative/highly cited papers—breaks the traditional global balance of science (Adams, 2013). The picture of the globalization of science as presented earlier is clearly linked to tensions in collaboration between the developed and developing (and richer/poorer in terms of gross domestic product [GDP] and higher education expenditure on research and development [HERD]) countries. Global network science opens incredible opportunities to new arrivals—countries as well as institutions and research teams. The advantages and disadvantages for the producers of traditional Euro-American top knowledge versus new entrants to global science collaborations differ, with possibly diversified implications for knowledge-producing personnel in developed/developing science systems. Globalization provides a context in which international collaboration in research provides channels through which developing countries can access the knowledge of developed countries more easily than ever before in the history of science. While, on the one hand, predominantly win-win collaboration types are certainly dominant (Wagner, 2008), free-riding behavior in knowledge production in developing economies is also possible, with possibly negative consequences for the global balance in the labor market for academic scientists (Freeman, 2010).

What is also at stake in the emergent tensions between the two clusters of countries (developed and developing) is public funding for academic research and the role of the public in the distribution of tax-based funding in the future. The core policy issue is why

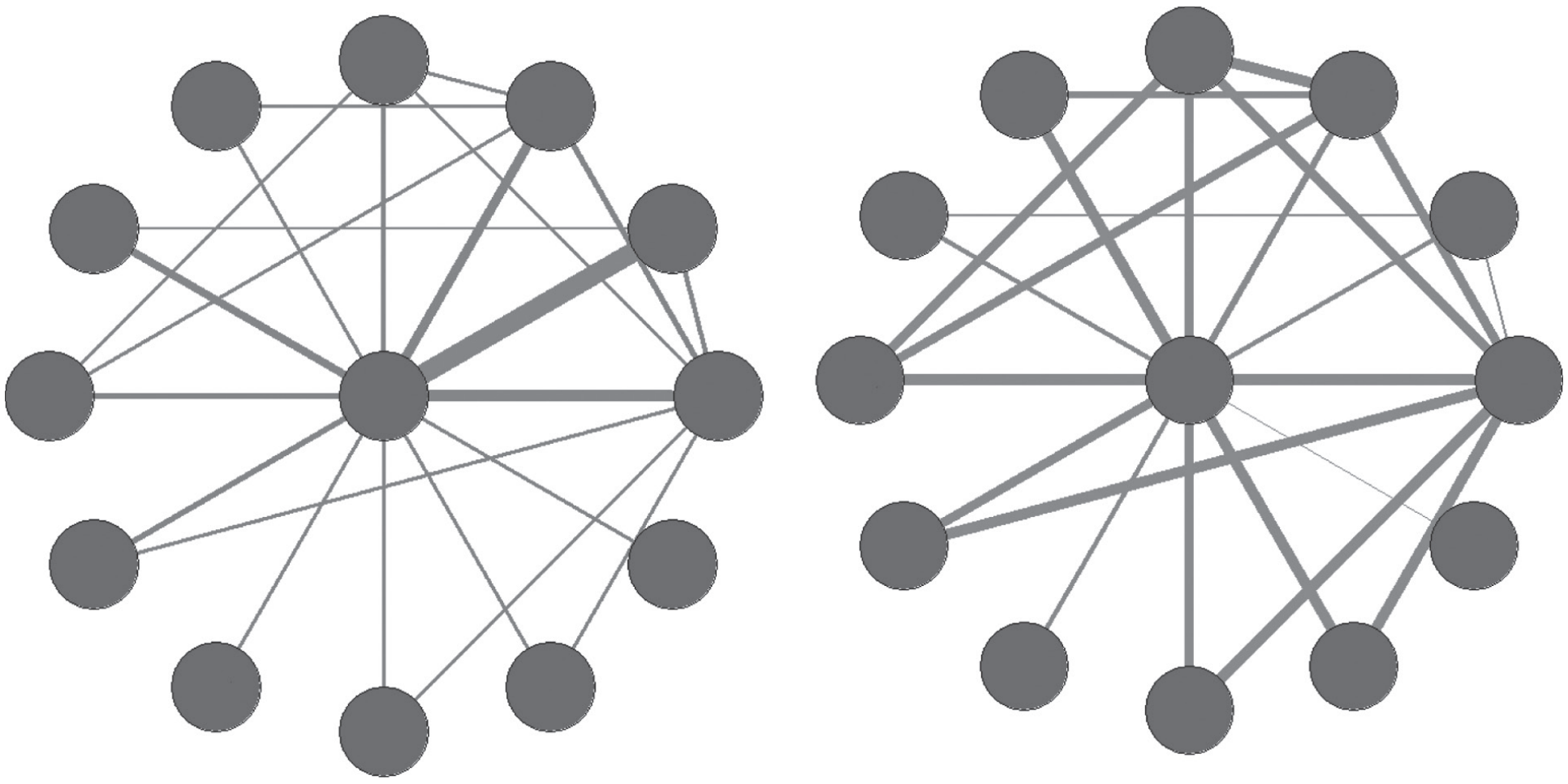

FIGURE 35.10 The network of internationally coauthored articles (in 2015–2020 combined), with only 25 most prolific pairs globally. Consequently, only edges with 59,148 (China-Canada) or more co-publications are displayed. The thickness of edges based on frequency of collaboration (left) and impact of collaboration as viewed through FWCI (right).

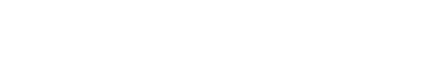

C35T4

**Table 35.4 Countries in This Chapter and Their ISO 3-Character Country Codes**

| | | | |
|---|---|---|---|
| AUS | Australia | ITA | Italy |
| BEL | Belgium | JPN | Japan |
| BRA | Brazil | KOR | South Korea |
| CAN | Canada | MEX | Mexico |
| CHN | China | MYS | Malaysia |
| CHE | Switzerland | NLD | Netherlands |
| CZE | Czech Republic | POL | Poland |
| DEU | Germany | RUS | Russia |
| ESP | Spain | SWE | Sweden |
| FRA | France | TUR | Turkey |
| GBR | United Kingdom | TWN | Taiwan |
| IND | India | USA | United States |
| IRN | Iran | | |

states fund academic research in general and fund highly internationally collaborative academic research as conducted in world-class universities in particular. The rationale presented by national governments may not fit the new reality of globally interconnected network science as conducted by highly internationalized scientists. Thus, national governments are indeed in a delicate position in which they seek national benefits and local applications in internationally produced and collaborative cutting-edge research, perhaps not being fully aware of the increasingly globalized and networked nature of science in which there appears to be no easy means to connect national funding to local benefits and applications. Policymakers and national funders of research may be immersed in the traditional vision of national science in contrast with individual scientists who are increasingly reaping the benefits of global science.

C35P56 The nature of the new global science fits perfectly into the always-present, more-private-than-public nature of implementing science for individual, career-enhancing purposes, with individual scientists and their motivations to conduct science at the very center of academic enterprise. Under the dominance of the Mertonian norms in the traditional imaginary of the academic profession, the role of this private nature has been systematically undervalued. However, we can trace the theme of the critical role of individual academic prestige and recognition in science in a long line of research from Hagstrom (1965) to Wagner (2018).

C35P57 The simple fact is that global science is funded by national governments; there is no global funding for research available on a large scale (except for philanthropy funding made available for selected grand challenges in research from global players such as the Bill & Melinda Gates Foundation, with total grant payments since inception totaling 54.8 billion USD in 135 countries). The national/global tension is much stronger in highly developed economies, with powerful academic science systems supported by strong public funding, than in less developed economies, with weak and publicly underfunded science systems. However, global science cannot be stopped, but the distribution of long-term gains and losses among collaborating partners in the global economy is far from clear, except for a general assumption that international research collaboration is good for global science and beneficial for societies at a global level, particularly from the perspective of science as a global collective good. However, in order to understand and apply knowledge and continue to function as full partners in global science, nations require their own science infrastructure and trained personnel, particularly doctoral students and young doctorates, even in difficult economic times (see Mattei, 2014). Consequently, as Chinchilla-Rodriguez et al. (2019, p. 6) argue, national scientific independence relies on government investment in research.

C35P58 There is a variety of forms of control in research collaboration, from loose to strict controls at various levels; in informal collaboration, typically, both governmental and institutional control are limited. However, governments, institutions, and funding agencies also have limited control over collaboration in the case of formalized and funded collaboration: the control over who collaborates with whom and who is doing what in science once funding is granted to national principal investigators with their international research teams is limited. The notions of international research collaboration as assumed by the funders, the grantees in national funding schemes, and their international collaborators may differ substantially. As Wagner (2006, p. 171) explained, "the question for developing countries is not how to get into collaboration with Germany, the United Kingdom or the United States, but how to take applicable knowledge from the network (no matter where it is located), make it relevant to local needs and problems, and tie it down." From the perspective of developing countries, the crucial aspect is to bring the results of collaboration back to the country, thereby enabling the meeting of local needs.

C35P59 The globalization of science does not imply that the global science system is created or planned by a single entity (the most natural candidate being the global science powerhouse, the United States); the global system is embedded in the rules created by scientists themselves and maintained as a self-organizing system. The implication of this is that nation states have another major level to consider in their science policies: the global level that accompanies, rather than replaces, the regional, national, and local levels. While public funding is key to the development of global science, innovations can occur anywhere and only scientists are able to identify and locate them and find ways to make them locally applicable. "It may become increasingly difficult to track spending outputs and outcomes, which has been the model for much of public accountability for science in the past" (Leydesdorff & Wagner, 2008, p. 324).

## C35S10 Global Science and the Power of Individual Scientists

C35P60 One thread continues across all previous sections and requires a summary: the rise of global science as closely related to transformations that occur at the shop-floor level of science and at the micro-level of individual scientists. Their motivation is important because collaboration choices at the micro-level of individual scientists determine international collaboration at the macro level of countries (Kato & Ando, 2017).

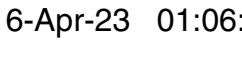

C35P61 There is substantial support in the literature regarding international research collaboration for the argument that the extent of such collaboration ultimately depends on the scientists themselves (Kato & Ando, 2016; King, 2011; Melin, 2000; Royal Society, 2011; Ulnicane, 2021; Wagner, 2008, 2018; Wagner & Leydesdorff, 2005). Faculty internationalization is considered to be shaped more by deeply ingrained individual values and predilections than by institutions and academic disciplines (Finkelstein et al., 2013) or, particularly, by governments and their agencies (Wagner, 2018).

C35P62 In their study on the role of global connectedness in the development of science in middle-income countries, Barnard et al. (2015) emphasized the increasing role of individual scientists. The global and national science systems are connected not so much through formal institutional collaborative ties but through individual scientists and their work: “it is the individual person which spans the local and the global worlds.” In other words, at the level of the individual researcher, there is no trade-off between local connectedness and global connectedness in research and they should be considered as “complements” but rather as “substitutes.” Consequently, the scientific connections between more and less advanced countries are created through individual scientists (Barnard et al., 2015, pp. 400–401).

C35P63 Perhaps, most importantly from the perspective of this chapter, the shift from a nationally centered scientific system to a global science system implies that it is increasingly the researchers, rather than national authorities, who set the rules of implementing science. The networked model of science is an open system, with opportunities open to new entrants, particularly new countries. However, it is individual scientists and their decisions that make the difference and change the course of science at the global level. Collaborative networks emerge from the choices of hundreds of scientists who shape the growth and evolution of networks “seeking to maximize their own welfare” (Wagner, 2008, p. 10).

C35P64 For decades, extant research literature has been dealing with the question of why academic scientists collaborate with other academic scientists. Perhaps the best answer is the simplest one: “scientists collaborate because they benefit from doing so” (Olechnicka et al., 2019, p. 45). From this perspective, scientists as “calculating individuals” are increasingly engaging in international collaborations because they are benefiting more from such collaboration—in terms of promotion, tenure, prestige, or access to research funding—rather than from any other type of collaboration (national, institutional). Scientists indicate “a pragmatic attitude to collaboration—when there is something to gain, then that particular collaboration will occur, otherwise it will not” (Melin, 2000, p. 39).

C35P65 Perhaps the most notable feature of science today is the presence of self-organizing networks, spanning the globe. These networks consist of researchers “who collaborate not because they are told to but because they want to . . . Scientific curiosity and ambition are the principal forces at work in the new invisible college” (Wagner, 2008, p. 2).

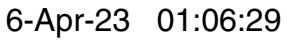

Scientists work within networks and the networks are constituted of the connections among these scientists. Scientists tend to collaborate across national borders because they “seek excellence” and want to work with the most outstanding scientists in their field (Royal Society, 2011, p. 57); they seek “resources and reputation” (Wagner & Leydesdorff, 2005, p. 1616); academic reward structures incentivize them to exploit collaboration and internationally coauthored publications to their own advantage (Glänzel, 2001). To this extent, collaboration is driven by an “intrinsic motivation to succeed” and “the motivation for better achievement” (Kato & Ando, 2016, p. 2). As such, it is largely curiosity-driven and reflects “the ambitions of individual scientists for reputation and recognition” (King, 2011, p. 24). The traditional postwar “governmental nationalism” in science coexists with this global science, as scientists believe that their curiosity-driven (rather than state-driven) approach “best serves their personal scientific ambitions” (King, 2011, p. 361).

C35P66 Wagner and Leydesdorff (2005, pp. 1610–1611) tested the hypothesis that global science is an emergent, self-organizing system where the selection of a research partner and research themes relies upon choices made by the scientists themselves. They tested whether international research collaboration could be shown to arise “from the self-interest of researchers to link together rewards, reputation, and the resources offered by a collaborative network,” referring to the concept of self-organization (see Melin, 2000; Ulnicane, 2021) and examining bibliometric data using network analysis. In addition, they studied the mechanism of preferential attachment at the field level and concluded that individual choices of scientists to collaborate internationally may be motivated by reward structures within science and influenced by the global abundance of collaborators and the weak ties among them: Weak ties are relatively easy to create and sever because people are not working side by side, and the social obligations that may arise from such collaboration within the same institutions are weaker.

C35P67 The relationship among major collaboration types—international, national, institutional, and solo research or no collaboration—are complex and depend on numerous factors that are internal or external to national science systems. The development of global networked science may be best viewed through preferential attachment mechanisms. Preferential attachment mechanisms employed to explain the individual behavior of scientists seeking collaboration imply that scientists wish to form links with other scientists of higher reputation or gain access to critical resources or funding: “preferential attachment clearly operates to the advantage of those at the top of the system, whether we think of them as individual scientists or as entire countries” (Wagner, 2008, p. 62). As Marginson comments, “researchers in the same or related disciplines want to work with each other. They fulfill their individual and collective agency by creating knowledge. . . . Knowledge flows freely, and science and its connections continue to grow and spread in all directions” (Marginson, 2020, p. 50). Therefore, the emergence of global science is indicative of the power wielded by individuals in science: “scientists and engineers are free to follow their own interests and careers wherever those may lead. . . . Most scientists will seek to enhance their reputations or gain access to resources, regardless of the interest of their nation of origin, and perhaps even at its expense" (Marginson, 2020, p. 64).

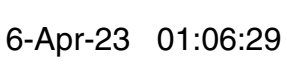

C35P68

Global science, regulated by intraprofessional interactions, provides agency to "autonomous researchers" (Marginson & Xu, 2021). Scientists rely on their "individual and collective goals, cognitive cultures, knowledge, imagination, associations, beliefs and habits" and global research agendas depend on global autonomous collegial networks (Marginson & Xu, 2021, p. 33). Marginson and Xu's notion of agency in collegial global science resonates well with the notion of free agents in global networked science in Wagner (2008) and the notion of autonomy in King (2011).

C35P69

As King (2011) emphasizes, the emergent global science enhances the opportunities for researchers to undertake collaborative projects across territorial boundaries that are beyond the direct control of national governments. Global networks in science are viewed as exceeding the power of governmental scientific nationalism, as they are privately governed and self-regulatory in nature. Scientists collaborate across the globe because the collaborative high-quality science satisfies their "individual curiosity and the career desire for esteem, reputation, and scientific autonomy" (King, 2011, pp. 370–371). Global science is controlled by researchers themselves, with key standardizing features being "stronger notions of autonomy, objectivity, testability, and peer judgment" (King, 2011, p. 372). The invisible college of global science is driven by the needs of the knowledge-creating community (Wagner, 2008, p. 32).

C35P70

In other words, global science provides more agency, autonomy, collegiality, and self-regulation to scientists embedded in national science structures and involved in global networks of science—unequal and highly stratified (Kwiek, 2019a, 2019b) but nevertheless open. The future of global science is in the hands of millions of scientists across the globe, who make individual decisions on whether or not to collaborate, and if collaborate—with whom, be they institutional, national, or international partners in research. Individual motivations drive scientists to collaborate in research and shape global science. It is safe to say that the role of individual scientists in the globalization of science (as well as the power of micro-level analysis in which individual scientists rather than institutions or countries are the unit of analysis) is underestimated and deserves much more scholarly attention.

C35P71

## Acknowledgments

C35P72

I gratefully acknowledge the research assistance provided by Dr. Wojciech Roszka. I also acknowledge the support of the Ministry of Science and Higher Education through its Dialogue grant 0022/DLG/2019/10 (Research Universities).

C35S11

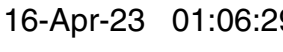

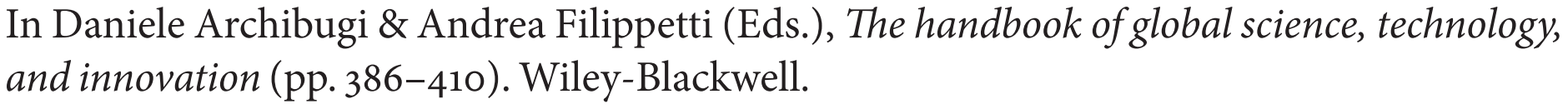

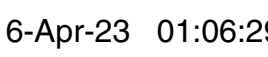

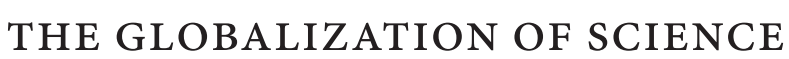